# Enhancement of seismic imaging: An innovative deep learning approach

Yanyan Zhang[*], Ping Lu[*], Hua Yu, Stan Morris

Anadarko Petroleum Corporation, Houston, TX, 77380

**Abstract**

Enhancing the frequency bandwidth of the seismic data is always the pursuance at the geophysical community. High resolution of seismic data provides the key resource to extract detailed stratigraphic knowledge. Here, a novel approach, based on deep learning model, is introduced by extracting reflections from well log data to broaden spectrum bandwidth of seismic data through boosting low and high frequencies. The corresponding improvement is observed from the enhancement of resolution of seismic data as well as elimination of sidelobe artifacts from seismic wavelets. During the training stage of deep learning model, geo-spatial information by taking consideration of multiple wells simultaneously is fully guaranteed, which assures that laterally and vertically geological information are constrained by and accurate away from the well controls during the inversion procedure. Extensive experiments prove that the enhanced seismic data is consistent with well log information, and honors rock property relationships defined from the wells at given locations. Uncertainty analysis could also be quantitatively assessed to determine the possibilities of a range of seismic responses by leveraging the outputs from the proposed approach.

**Introduction**

Seismic data bandwidth extension has been widely investigated for resolution enhancement purpose. In fact, seismic wave is frequency dependent and it gets attenuated through the process of propagation, such that the higher frequency signal is absorbed more rapidly compared to lower frequency (Ma, Zhu, Guo, Rebec, & Azbel, 2005). The seismic resolution defines its capability to separate thin beds and with high resolution seismic data, the reflections from the top to bottom of a bed can be detected accurately.

Most of the existing research have been focusing on extending the limit of seismic high frequency in spectrum (Stark, 2009). On the other hand, low frequency components are also desirable for seismic inversion purpose. In this work, we purpose a deep-learning based technology to not only boost the high end but also extend more low frequency parts in seismic spectrum.

---

[*] These authors contributed equally to this work.

**Using deep learning model to directly extend seismic frequency bandwidth**

Broadband seismic data is in demand as it provides more details about deposition from high resolution. However, due to the nature of seismic acquisition and processing, it is hard for existing techniques to extend the frequency while maintain an acceptable signal-to-noise (SNR) ratio, which implies the majority of current seismic bandwidth extension methods are done by adding more noise to seismic traces (Liang, Castagna, & Torres, 2017).

Instead, the purposed deep learning architecture learns the distribution and correlation between well log signal and the seismic trace in this location. This is a supervised learning model, and well log information is utilized as an extended bandwidth trace with both low and high frequency components. During training, the network gets seismic trace as input and well log signal as output to learn the transformation in between, and then in inference process the network is filled by all the seismic traces in testing volume as input and let the generative model to predict the synthetic broad frequency signal correspondingly.

One crucial part regarding of the purposed training structure is seismic well tie. Ideally, if seismic is a robust estimation of bandlimited reflectivity, synthetic filtered well log can tie up with seismic and we can utilize the original log as broad frequency version of seismic. However, due to the unavoidable attenuation in seismic acquisition, it is not always reliable to reflect structures in subsurface. There are two standards, amplitude and characterization, to evaluate the well log seismic tying. Deep learning model is less sensitive to amplitude-wise untie, however, when suffering bad seismic well tie in characterization, low frequency seismic may have no or even opposite reflection compared to log, and we should avoid using such pair of data as training input to make sure the network is unconfused. In our dataset, we have twelve seismic log pairs in total and nine of them have good tie, two of them have fair tie and one has poor tie in characterization. To leverage more possibilities to the network in training, we picked three pairs of good tie and one pair of fair tie data as training input and left the rest of the pairs for validation purpose.

**Deep Learning Architecture: Conditional Generative Adversarial Networks**

Motivated by the great success that deep learning architectures, especially generative adversarial networks (GANs) (Lu, Morris, Brazell, Comiskey, & Yuan, 2018) have achieved in seismic interpretation tasks, we purpose a conditional GAN based deep learning network to fulfill the objective of seismic bandwidth extension.

Traditional deep convolutional GANs (Goodfellow, et al., 2014) (Radford & Metz, 2015) consists of two components, a generator used to generate new, synthetic instance and a discriminator used to distinguish whether each instance is from the actual dataset or from the generator. For DCGANs, the generator has no information about the actual dataset and it learns to create synthetic images by interacting with the discriminator. However, as the input of generator is purely random noise vector, there is no control over the particular type of images that would be produced. To overcome such limitation, conditional GANs (Mirza & Osindero, 2014) (Isola, Zhu, Zhou, & Efros, 2017) has been purposed by allowing additional information to be accessible to both generator and discriminator, such that the discriminator can penalize generator if the structure of synthetic image does not match with the condition information.

Specifically for our task, during training process, the spectrogram of each seismic trace is served as the condition information observable to both generator and discriminator, and the spectrogram of corresponding well log trace is used as actual image to train discriminator. Both generator and discriminator use the structure of module as convolution layer, followed by Batch Normalization and ReLu activation layers to extract the image features. A detailed flow chart for purposed network is displayed at Figure 1. The noise vector inputted at generator side along with the condition image is to overcome overfitting issue.

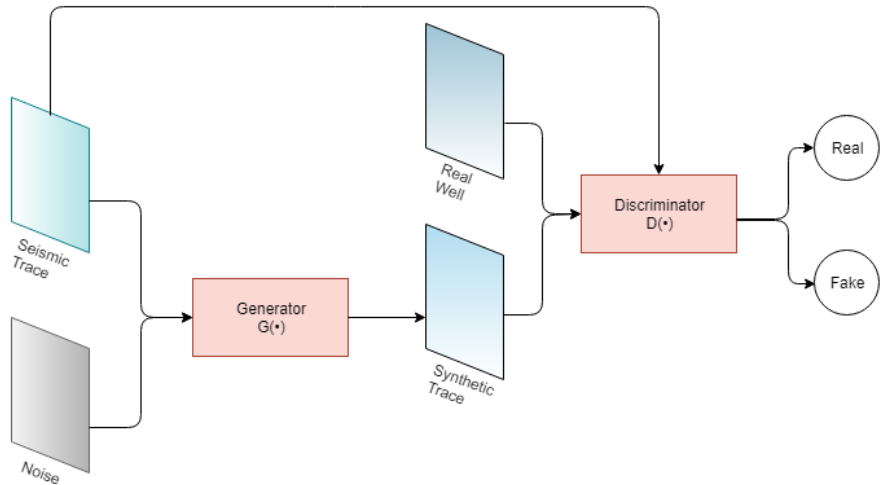

Figure 1: Conditional GANs flow chart

The objective function of conditional GANs architecture is

$$\min_G \max_D E_{(x,y)}[\log D(x,y)] + E_{(x,z)}[\log(1 - D(x, G(x,z)))], \quad (1)$$

where the discriminator $D(\cdot)$ tries to maximize such loss while the generator $G(\cdot)$ is against it. Inside the equation, $x$ denotes the conditioned input image (seismic trace), $y$ is the desirable output of the network (well log information) and $z$ is the Gaussian random noise.

In addition, since the proposed learning framework is a supervised learning and we require the generated synthetic trace to be similar to the ground truth well log trace, a regularization term is necessary to control the generated synthetic image as

$$E_{(x,y,z)}[\|y - G(x,z)\|_1]. \quad (2)$$

Here, $L_1$ norm is selected to avoid potential blurriness caused by $L_2$ distance.

From the objective function above, it can be seen the purposed network focuses on learning the structure loss, which penalizes not individual pixel but the joint configuration of the output image. Such setup motivates the structure of output to be aligned with the structure of input, so that the geology from original seismic remains unchanged while the generator adds more frequency contents into the traces.

**Statistical Analysis**

Generative models introduce non-controllable randomness into the workflow, whereas the objective of performing bandwidth extension is to achieve more detailed seismic with high accuracy. Therefore, statistical analysis is required to validate whether the output from the proposed DL model could converge to reflect the precise seismic events.

Comprehensive experiments have been done to check the distribution of multiple realizations for the same input. Specifically, we select several seismic traces in the field such that each seismic trace contains 512 samples in depth, and we collect 500 different realizations from various training epoch for each sample. Histograms for randomly selected sample points in depth are displayed in Figure 2, which shows that all the 500 realizations are highly centralized with minor variation, and it implies the mean value of multiple outputs would be a good measure for true result.

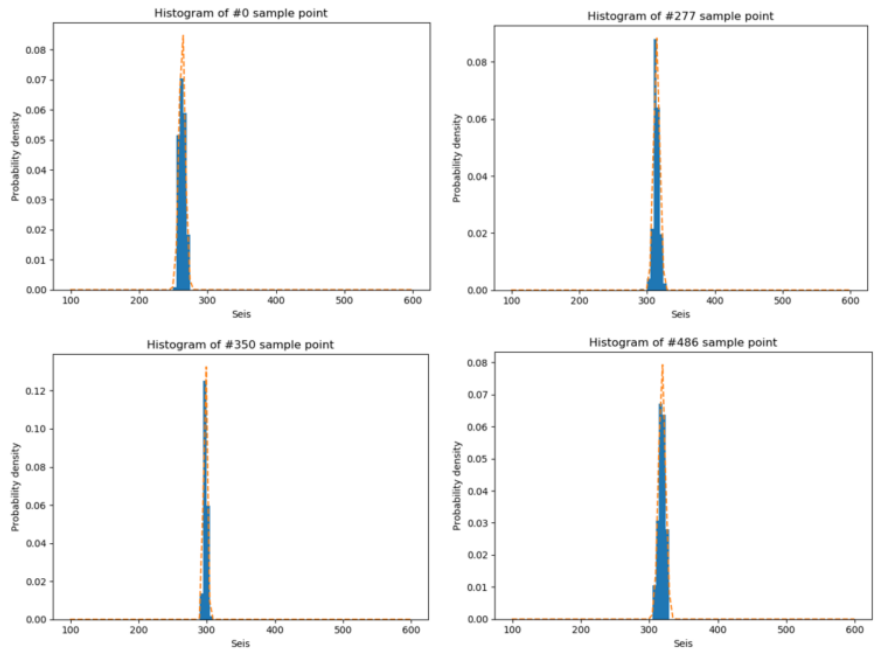

*Figure 2: Histograms from 500 realizations for randomly selected seismic sample points in depth.*

In the following discussion, all the generated broad frequency seismic traces are based on averaging multiple realizations for each individual sample point.

**Further Discussion s on Various Training Combinations**

As discussed, cGANs architecture applied to the problem is considered supervised learning, which requires the selection of training dataset. There are two major aspects we considered during the selection process: seismic-well tie and the type of lithology in log data. If the training seismic-well ties poorly, the network may output opposite characterizations in broadband trace during inference. On the other hand, if all the training input only contain one type of lithology, blocky sand for example, then the network may lose the capability to generate thin sands, as it has no information for the existence of other types of facies. In either case, the proposed DL network suffers from unsatisfactory training information, which could potentially lead to produce inaccurate broadband seismic volume. Figure 3 shows outputs from different training combinations for same validation well location, which have non-trivial distinctions in between. Based on these facts, we encourage researchers to investigate seismic-well tie and facies distribution in targeted region before choosing the training information. Our philosophy is to leverage more possibilities to the DL network, so we pick training pairs from multiple facies and good or fair seismic-well tie. A more rigorous discussion on training information selection could be treated as a future work in the field.

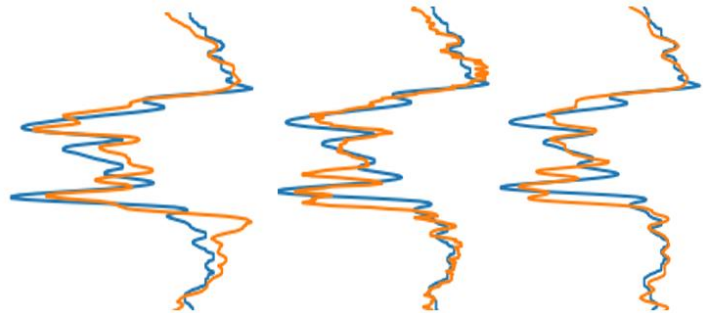

Figure 3: Real well log (blue curve) vs. synthetic seismic trace (orange curve) for same well location from different training combinations.

It is also worth to mention that the number of training pairs is a hyper-parameter and could be decided by having multiple experiments. Although in this work we use 4 pairs for training input, the network is not limited to this number and could be fed in any size depends on specific dataset.

**Experiment on Field Offshore Data**

The proposed DL based bandwidth extension technique has been applied to an offshore dataset to prove the concept, in which multiple seismic-well pairs are available for training and validation purposes. The DL architecture is trained with 4 pairs of data, each contains one well log and its corresponding seismic trace at this location. Figure 4 shows comparison between real logs versus generated broad bandwidth traces for both training and validation information. The first two signals at Figure 4A shows the original log (red) and the log filtered to the detachable frequency band (blue) used for training, and Figure 4B shows the real seismic trace at this location (red) and the DL predicted log (blue) generated from it. It can be seen that for the training pair, the synthetic broad bandwidth seismic matches the log almost perfect, such that both detailed spikes and curve are captured. The right image shows comparison between blind well and synthetic broad-frequency seismic generated by DL network. In each image, blue curve is the real log and yellow curve is the synthetic trace. For these validation pairs, synthetic broad frequency seismic still follows the signal variation and matches the majority of log spikes.

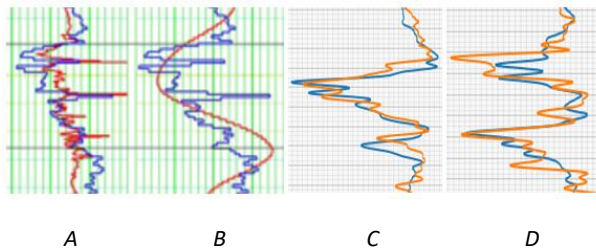

A   B   C   D

Figure 4: Comparison between real logs versus generated broad bandwidth traces for training (left) and validation (right) information.

To take a closer look, Figure 5 illustrates the wavelet and spectrum comparison between the original seismic data (blue) and DL-synthetic prediction (green). The wavelet shape of the DL-synthetic results is more centered and has much less sidelobe than the original seismic data. In addition, the spectrum analysis of the original seismic and DL-prediction shows that the DL network not only boosts the high frequency components of the seismic, it also generates more low frequency information to the synthetic trace, which proves the claim that the proposed DL network is capable of extending a broader bandwidth for seismic volume.

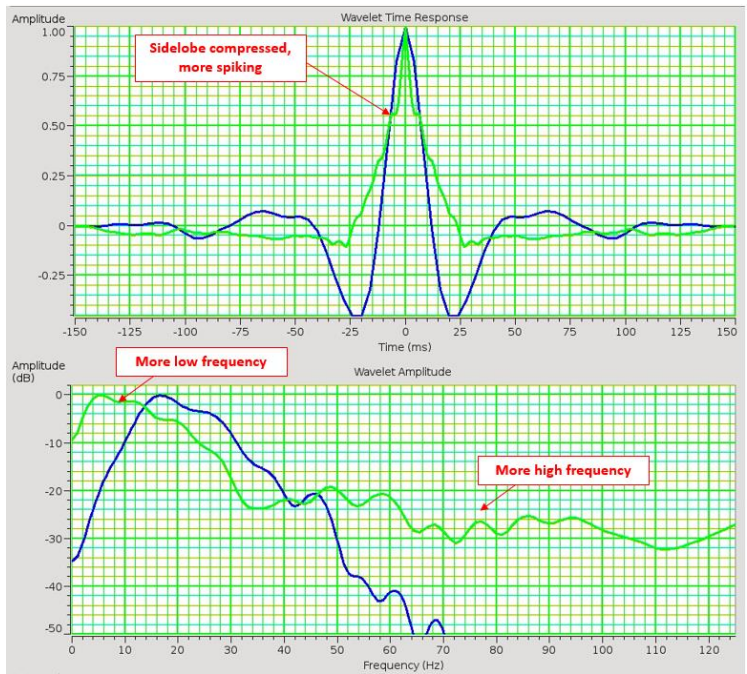

Figure 5: Comparison between seismic trace and its corresponding synthetic broad frequency signal outputted from DL network, with spectrum analysis displayed below.

On top of the success with trace validation, experiments have been done for a 3D seismic volume to demonstrate the effectiveness of the method. We input all seismic traces from an asset and collect the outputs as generated broadband volume. Figure 6-9 show the comparison between original seismic data and results from DL network with different filter schemes. In Figure 6-9, A is the gamma ray log from well, B is the input original seismic data, far angle stack with minus 90 phase rotation, and overlain by corresponding elastic impedance. C is the DL-predicted result.

In Figure 6, the original seismic used as input for DL network is compared to the original image of synthetic band extended seismic without any filtering. Due to limited resolution and tuning effect, original seismic image suffers from the sidelobe and detects much thinner sand body against the log. However, with our produced broadband image, the sidelobe of seismic successfully get removed, and the predicted sand thickness honors the log information.

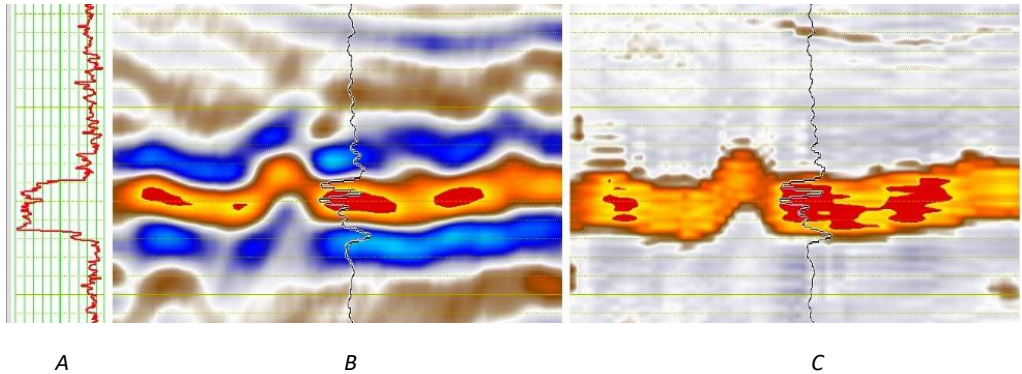

*A                              B                                C*

*Figure 6: Original seismic versus unfiltered outputs from DL network.*

However, as the high frequency components may contain more noise with less information, Figure 7 shows the seismic image confined with 0-50-250-500 band, which targets on the most interested frequency region. Although trivial sidelobe is introduced due to the filtering, the detected sand thickness remains accurate and the synthetic broad frequency seismic reflects more thin layers with granularities.

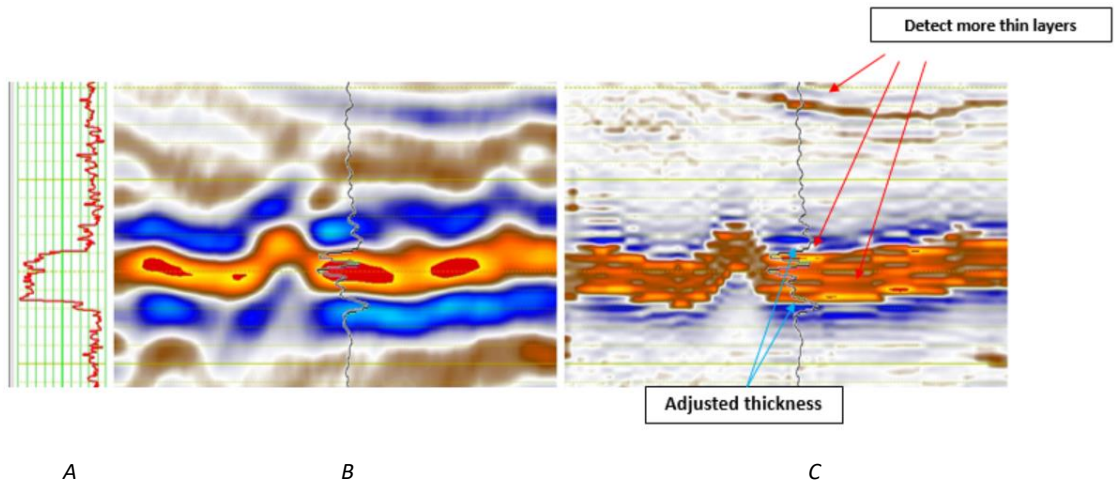

*A                              B                                C*

*Figure 7: Original seismic versus 0-50-250-500 band limited outputs from DL network.*

In order to validate the whole DL process, we filtered the DL-predicted results back to seismic frequency range (3-6-60-80). The results are shown in Figure 8. Even though the generated seismic has lower resolution than before, it still presents a thicker sand body matching the information from well log, as well as weakens the sidelobe effects on seismic image.

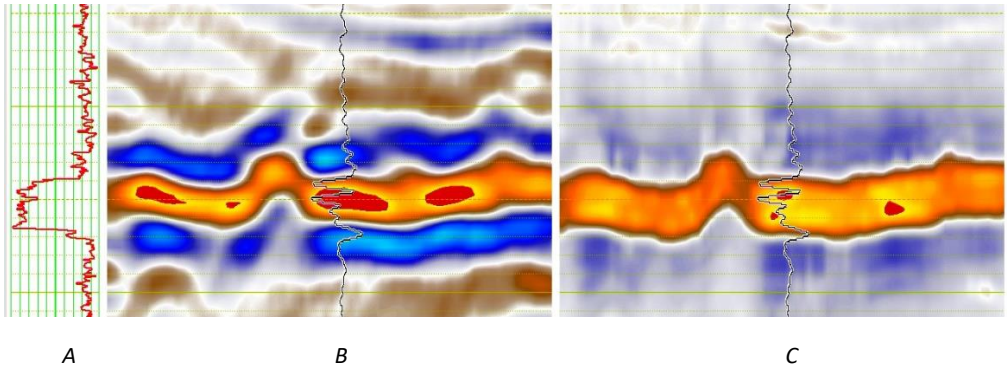

A               B               C

*Figure 8: Original seismic versus seismic bandwidth (3-6-60-80) outputs from DL network.*

As our method can not only boost the high frequency part but also add more information to the low frequency region, we confine the generated seismic with filter 0-0-8-16 to focus only on the low frequency details. It can be seen in Figure 9 that, compared to the original seismic, strong reflection in the middle of the low frequency image indicates continuous sand body, which also honors the characteristics of the log.

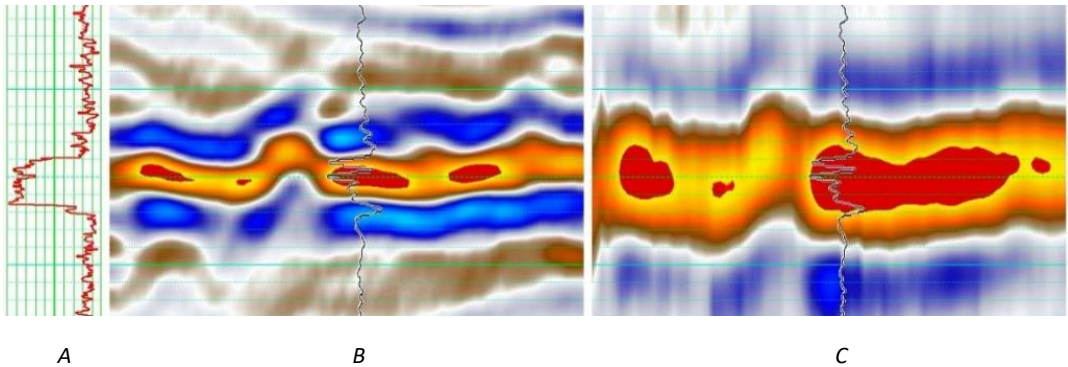

A               B               C

*Figure 9: Original seismic versus low frequency bandwidth (0-0-8-16) outputs from DL network.*

Lastly, Figure 10 displays the difference between original seismic and the generated outputs from our proposed DL architecture in depth slice. Our broad bandwidth seismic maintains the geology features from input information, such as the faults in original image remain unchanged. In the meanwhile, the generated broadband seismic image has high resolution, revealing more detailed deposition information and clear boundaries between various facies.

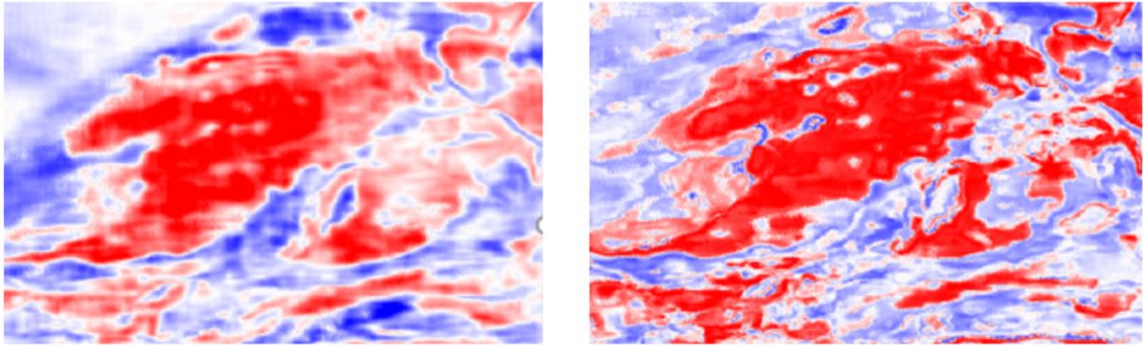

*Figure 10: Original seismic versus broadband seismic from DL network in depth slice.*

**Conclusions**

The purposed deep learning network can boost the seismic frequency to the extent of log scale in both low and high ends, while removing sidelobe and creating more spiked amplitude. As the resolution of seismic image has been improved, the generated broad frequency seismic data can reveal more thin layers and adjust the sand bed thickness honoring the log information. In addition, the broadband seismic matches perfectly in the training log location while it also achieves satisfactory similarity in the blind well location, which further validates the effectiveness of our proposed method.

**Acknowledgement**

The authors thank Anadarko for permission to publish this work.